\documentclass[11pt]{article} 

\usepackage[utf8]{inputenc}
\usepackage[english]{babel}
\usepackage{graphicx}
\graphicspath{ {images/} }
\usepackage[a4paper,right=1in,left=1in,top=1in,bottom=1in]{geometry}
\usepackage{fancyhdr}
\usepackage{rotating}
\usepackage{placeins}

\usepackage[round]{natbib}

\usepackage[dvipsnames]{xcolor}
\definecolor{light-gray}{gray}{0.95}
\colorlet{laurencol}{cyan!10!light-gray}
\colorlet{laurencolvert}{cyan!25!light-gray}
\colorlet{laurencol2}{red!10!light-gray}
\usepackage{colortbl}
\usepackage{arydshln}

\makeatletter
\newcommand\notsotiny{\@setfontsize\notsotiny\@vipt\@viipt}
\makeatother

\usepackage{tikz}

\usepackage{tikz}
\usetikzlibrary{shapes,arrows,decorations.markings}
\usetikzlibrary{positioning}

\usepackage{amsmath, amsfonts, amssymb} 
\usepackage[labelfont=bf]{caption}
\usepackage{setspace}
\usepackage{accents} 
\usepackage{color, soul}
\usepackage[title,titletoc,toc]{appendix}
\usepackage{rotating} 
\usepackage{float} 
\usepackage{multirow} 
\usepackage[position=top]{subfig}
\usepackage[subfigure]{tocloft}
\usepackage{tikz}
\usetikzlibrary{decorations.pathreplacing,calc}
\usepackage[many]{tcolorbox}
\newtcolorbox{rightbrace}{%
    enhanced jigsaw, 
    breakable, 
    frame hidden, 
    parbox=false,
}
\usepackage[explicit]{titlesec}

\allowdisplaybreaks
\usepackage{color}   
\usepackage{hyperref}
\hypersetup{
    colorlinks=true, 
    linktoc=all,     
    linkcolor=black,  
    filecolor=black,
    citecolor = black,      
    urlcolor=black,
}

\definecolor{laurencol0}{rgb}{.1,.6,.3}
\colorlet{laurencol}{laurencol0!20!white}

\title{Addressing delayed case reporting in infectious disease forecast modeling }
\author{\textbf{Lauren J. Beesley$^{*1}$, Dave Osthus$^{1}$, and Sara Y. Del Valle$^{1}$} \\
$^{1}$Los Alamos National Laboratory, Los Alamos, NM\\
*Corresponding Author: lvandervort@lanl.gov
}
\date{}

\begin{document}
\maketitle 

\abovedisplayskip=6pt
\belowdisplayskip=6pt
\allowdisplaybreaks
\raggedbottom

\begin{abstract}
Infectious disease forecasting is of great interest to the public health community and policymakers, since forecasts can provide insight into disease dynamics in the near future and inform interventions. Due to delays in case reporting, however, forecasting models may often \textit{underestimate} the current and future disease burden. \\
\indent In this paper, we propose a general framework for addressing reporting delay in disease forecasting efforts with the goal of improving forecasts. We propose strategies for leveraging either historical data on case reporting or external internet-based data to estimate the amount of reporting error. We then describe several approaches for adapting general forecasting pipelines to account for under- or over-reporting of cases. We apply these methods to address reporting delay in data on dengue fever cases in Puerto Rico from 1990 to 2009 and to reports of influenza-like illness (ILI) in the United States between 2010 and 2019. Through a simulation study, we compare method performance and evaluate robustness to assumption violations. Our results show that forecasting accuracy and prediction coverage almost always increase when correction methods are implemented to address reporting delay. Some of these methods required knowledge about the reporting error or high quality external data, which may not always be available. Provided alternatives include excluding recently-reported data and performing sensitivity analysis. This work provides intuition and guidance for handling delay in disease case reporting and may serve as a useful resource to inform practical infectious disease forecasting efforts.
\end{abstract}

\section{Introduction} \label{intro}
Accurate model-based forecasts of future disease rates are essential for informing public health policy and response to infectious disease outbreaks. Strategies for improving forecasts are of great interest to the public health and epidemic modeling communities, and use of the most up-to-date reports of case diagnoses can often provide useful insight into disease dynamics in the near future. For many diseases, however, surveillance and case identification take time, and there is often a delay between when people become \textit{infected} with an infectious disease and when that infection is officially \textit{reported} and made publicly available \citep{Jajosky2004}. \\
\indent Often, reporting at the national level requires many intermediate reporting steps; diagnostic labs and health care systems report suspected and/or confirmed cases to local public health systems, which then report to states, which then report to the national level \citep{Jajosky2004}. Delays in each one of these reporting steps can result in substantial lag in the information available for forecasting, and these delays may vary on a variety of factors, including state infrastructure and threshold for reporting suspected vs. confirmed cases. National, state, and local reporting systems  produce regular reports providing the most up-to-date statistics on disease diagnoses (e.g., weekly statistics provided by the National Notifiable Diseases Surveillance System), and these provisional reports may be revised/corrected through a process called backfill as new data become available. Real-time reports of current disease burden often substantially underestimate the number of cases eventually reported (hereafter referred to as the validation case counts) for a given time period \citep{McGough2020}. \\
\indent When the goal is to forecast future disease burden, it is not always clear how one should use these error-prone real-time case reports. When the reporting delay is minimal, we may expect reporting to have little impact on model forecasts and use these real-time reports without correction. When reporting delay is substantial and when accurate real-time estimates and forecasts of disease burden are needed to inform policy interventions, additional thought is needed to address the reporting delay. \\
\indent One common strategy for handling reporting delay is to rescale real-time case reports using the expected amount of under-reporting \citep[e.g.,][]{Lawless1994}. Calculating the scaling factors (hereafter referred to as reporting factors) presents a complication, and more sophisticated strategies have been used to model them and the disease process jointly \citep[e.g.,][]{Hohle2014, McGough2020, Bastos2019, England2002}. Other work has used external data to predict the validation cases (e.g., social media, Google Trends) and account for reporting delay using a weighted version of the real-time data and the validation data predictions \citep{Osthus2019b}. Backcast imputation approaches handle the reporting error by generating past weeks' validation data using the most recent data reports for those weeks \citep{Brooks2018}. These methods have been studied across multiple disciplines beyond infectious disease modeling such as in actuarial claims prediction \citep{England2002} and in correction of cancer registry case reporting \citep{Midthune2005}. Some of these studies compare two or three methods for addressing reporting delay head to head in narrow modeling settings, but there are no recommendations for how reporting delay should be handled for infectious disease forecasting in general practice. \\
\indent In this paper, we propose a general framework for addressing reporting delay in terms of a two-stage estimation problem: (1) estimation of the reporting factors as a function of time since the initial report and (2) disease modeling and forecasting, accounting for under- or over-reporting. We propose several novel strategies for leveraging either historical data on case reporting or external social media/Google trends data to estimate the amount of reporting error. We then describe how existing strategies for addressing reporting delay can be implemented for infectious disease forecast modeling in general. We apply these methods to address reporting delay in data on dengue fever cases in Puerto Rico from 1990 to 2009 and to reports of influenza-like illness (ILI) in the United States between 2010 and 2019. Through a simulation study, we compare how these methods perform when all assumptions are satisfied and evaluate their robustness when assumptions are violated. This work provides intuition and guidance for the comparative performance of various methods for handling delay in disease case reporting and may serve as a useful resource to inform practical infectious disease forecasting efforts.\\
\indent In \textbf{Section \ref{app}}, we introduce the data examples, and we describe the various conceptual approaches for handling reporting delay in \textbf{Section \ref{methods}}. We implement various approaches for several data examples in \textbf{Section \ref{apppres}}, and we explore performance and robustness of these methods in greater detail in \textbf{Section \ref{sims}}. We present a discussion in \textbf{Section \ref{discuss}}.

\section{Disease surveillance data with reporting delay}  \label{app}
We consider two motivating datasets in which delayed reporting of disease cases results in discrepancies between the number of cases reported in real-time and the number of cases eventually reported. In our first data example, we model cases of dengue fever reported between 1990 and 2009 in Puerto Rico. In our second data example, we explore influenza-like illnesses in the United States between 2010 and 2019.

\subsection{Dengue fever surveillance data in Puerto Rico, 1990-2009}  \label{dengue_intro}
\begin{figure}[h!]
  \centering
\caption{Proportion of eventually-reported cases that were reported in each week $^1$ }
\subfloat[Puerto Rico dengue fever]{\includegraphics[trim={0cm 0cm 0cm 1.2cm}, clip, width=3in]{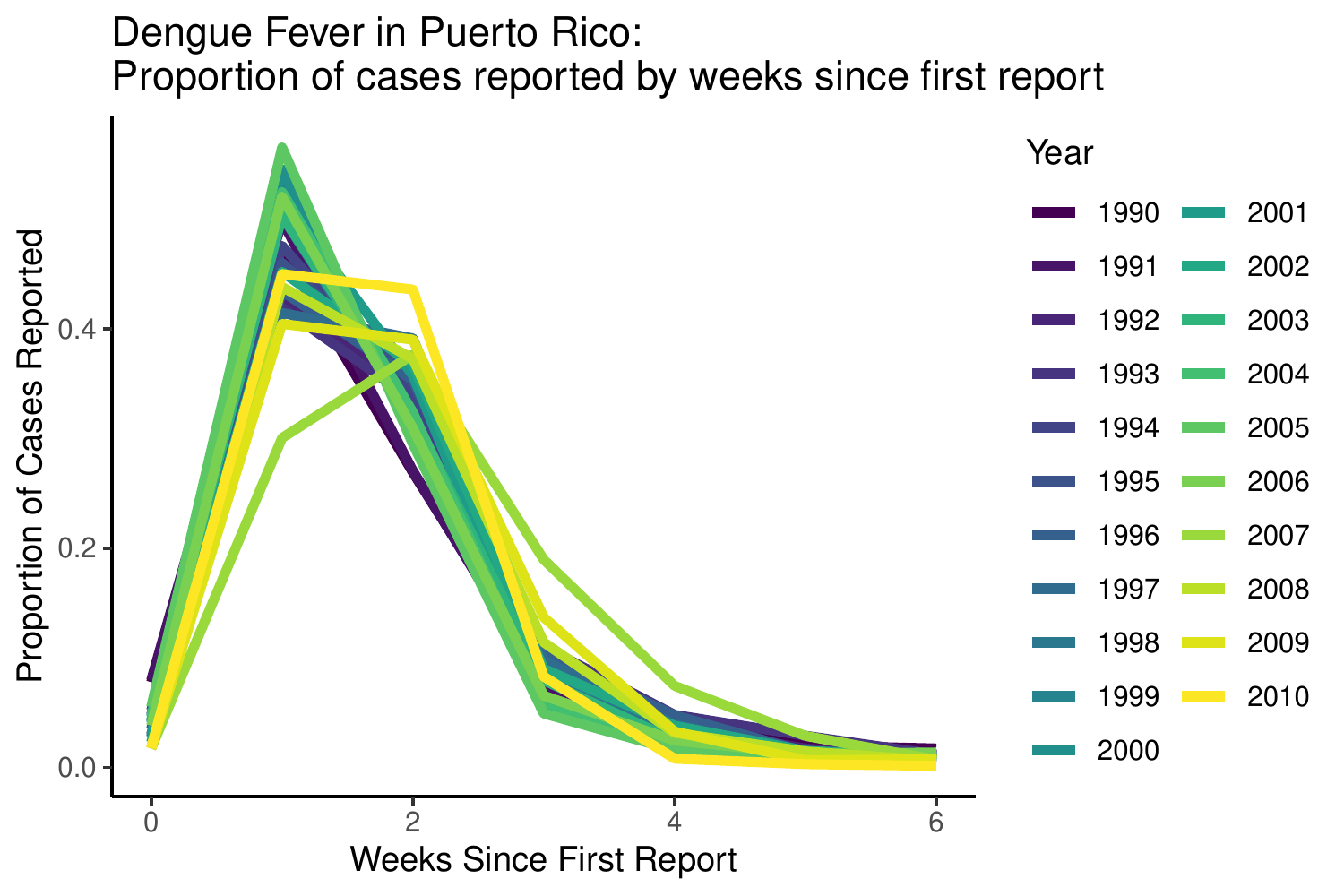}}
\subfloat[National US Influenza-like Illness]{\includegraphics[trim={0cm 0cm 0cm 1.2cm}, clip, width=3in]{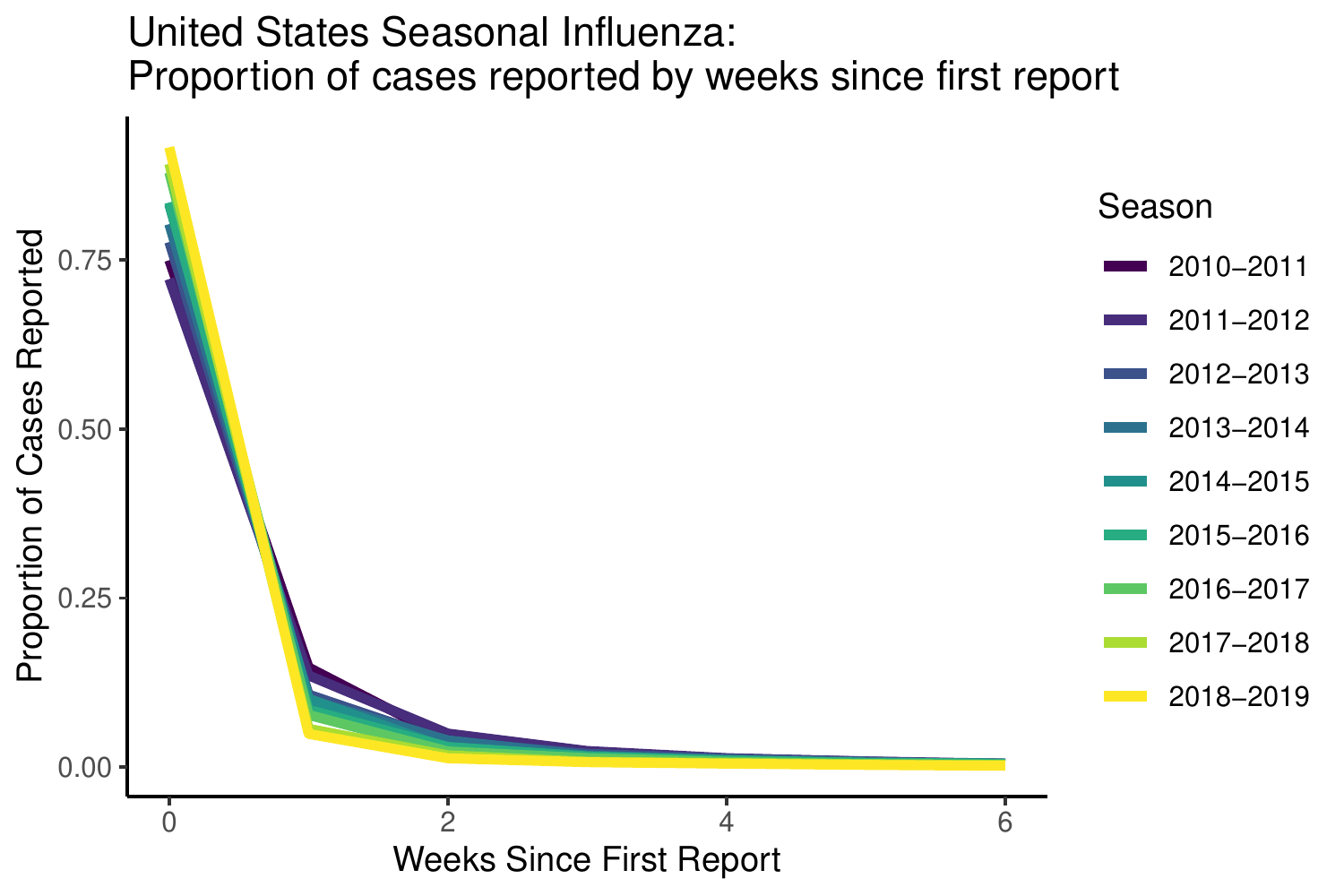}}
\caption*{ \footnotesize  $^1$ Estimates of $\pi_{ts}(d)-\pi_{ts}(d-1)$, obtained using \ref{preest} and stratified by season/year }
\label{pi_viz}
\end{figure}

We obtain publicly-available data on dengue fever cases in Puerto Rico between 1990 and 2009 that are provided alongside \citet{McGough2020} as part of the R package \textit{NobBS}. These data provide the official diagnosis week and the week of case reporting for over 53,000 cases of laboratory-confirmed dengue fever in Puerto Rico as collected by the Puerto Rico Department of Health and the Centers for Disease Control and Prevention. Using these person-level data, we calculate the number of cases \textit{diagnosed} in each calendar week, which we call the validation data. For each calendar week, we also calculate the number of these validation cases that were actually \textit{reported} as part of the initial case counts from the first week and the number that were reported in each of the following 6 weeks. These weekly reporting data were constructed to mimic weekly ``official" case reports. We will refer to these constructed ``reports" throughout, but it should be understood that these are hypothetical weekly reports rather than officially-released governmental statistics. A subset of the resulting case counts are shown in \textbf{Supp. Figure A1}. In \textbf{Figure \ref{pi_viz}a}, we show that only a very small proportion ($\sim 5\%$) of eventually-reported cases were included in the initial case ``report" (i.e., officially reported within 1 week of diagnosis). The vast majority of cases were reported in the following six weeks. 

\subsection{Seasonal influenza-like illness in the United States, 2010-2019}  \label{influenza_intro}
We also consider publicly-available data on influenza-like illness (ILI) in the United States between 2010 and 2019 compiled as part of the U.S. Outpatient Influenza-like Illness Surveillance Network (ILINet). Both validation and officially-reported real-time data (i.e., data that were available up to $d$ weeks after the initial case report) are available at https://github.com/cmu-delphi/delphi-epidata using the Delphi API \citep{Farrow2015}. Validation data were defined as the case counts for each calendar week that were reported as of June 13th, 2021. In modeling of ILI cases, it is common to break up the data recorded by calendar year into flu seasons, where weeks are re-defined relative to the start of the season. For our analysis, we identified the start of each 35-week flu season as the 40th week in the calendar year as defined by the Morbidity and Mortality Weekly Report (i.e., MMWR week 40) produced by the US National Notifiable Diseases Surveillance System \citep{NNDSS}. For example, the 2018-2019 \textit{season} spans the end of 2018 and the beginning of 2019. In addition to the total number if ILI cases nationally, we also downloaded data stratified by state from 2010 through the 2018-2019 season. Real-time reporting data were available for the 2017-2018 and 2018-2019 seasons and also for the 2016-2017 season for some states. Florida was excluded from analysis due to the lack of real-time reporting data provided by the Delphi API. ILI cases occurring after the 2018-2019 season were not included in this analysis to avoid inclusion of COVID-19 diagnoses.\\
\indent A mild amount of case reporting delay was evident for these data, where roughly 70-90\% of eventually-reported cases were reported initially (\textbf{Figure \ref{pi_viz}b}). There was a clear seasonal trend in reporting at the \textit{national level}, where more recent seasons had a higher proportion of validation cases reported initially. Unlike the dengue fever data, we did not obtain disease diagnosis and reporting dates for individual patients; rather, we considered the total number of reported cases for each week. Due to excluded providers or other reasons, it was possible for real-time case counts to be larger than the validation counts. Therefore, the discrepancy between the real-time reported and validation case numbers was due to a combination of under-reporting (i.e., not-yet-reported cases) and over-reporting (cases currently reported that were eventually excluded). This phenomenon was particularly evident in the 2017-2018 season data for Vermont, where the number of reported cases often decreased as the number of weeks since the initial report increased (\textbf{Supp. Figures A2-A3}).


\section{Methods }  \label{methods}
\begin{figure}[h!]
  \centering
\caption{Summary of Methods for Handling Reporting Delay}
\subfloat[Method details]{\includegraphics[trim={0cm 0cm 0cm 0cm}, clip, width=6in]{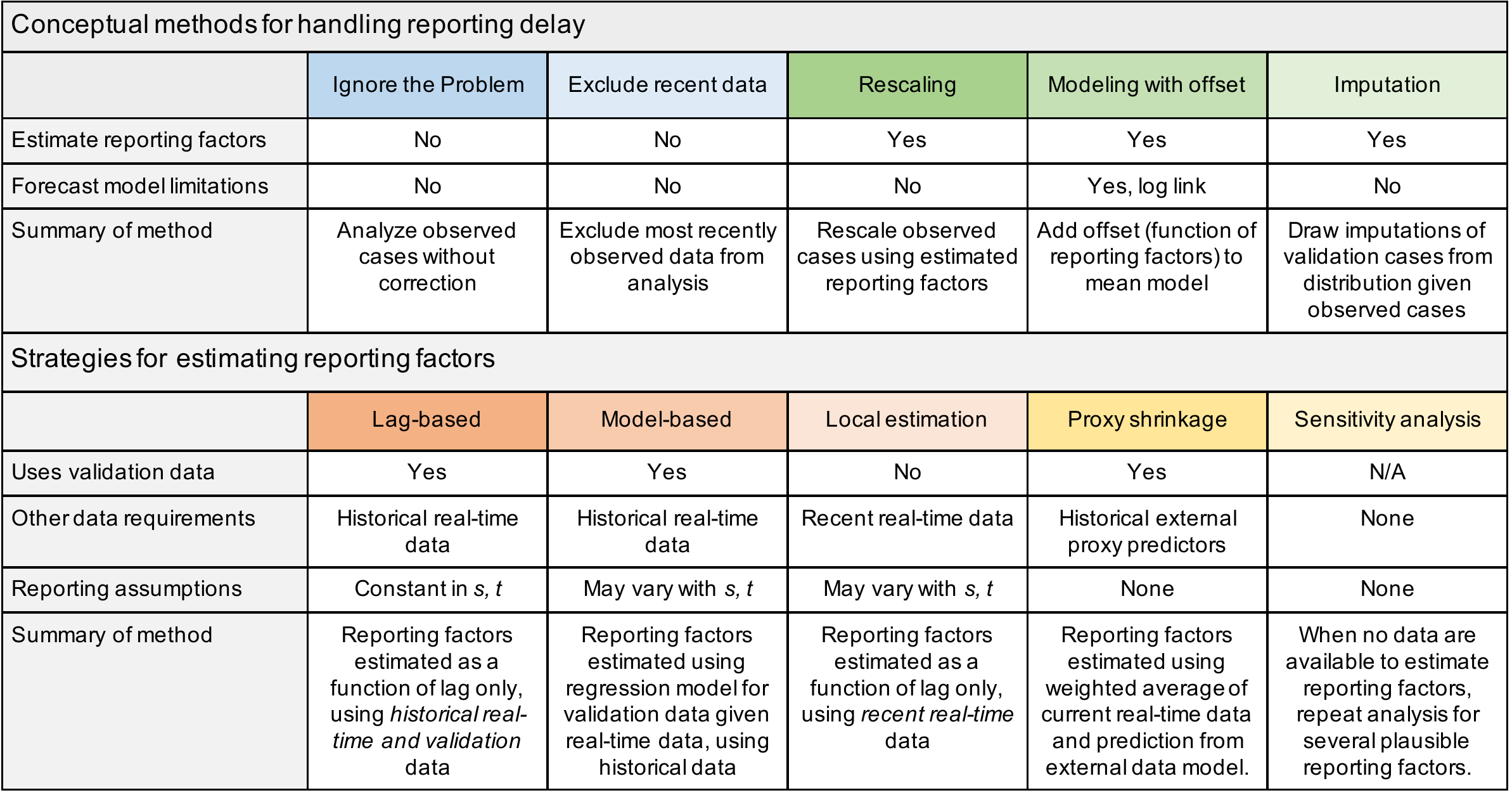}}\\
\subfloat[Flowchart of method recommendations]{\includegraphics[trim={0cm 0cm 0cm 0cm}, clip, width=6in]{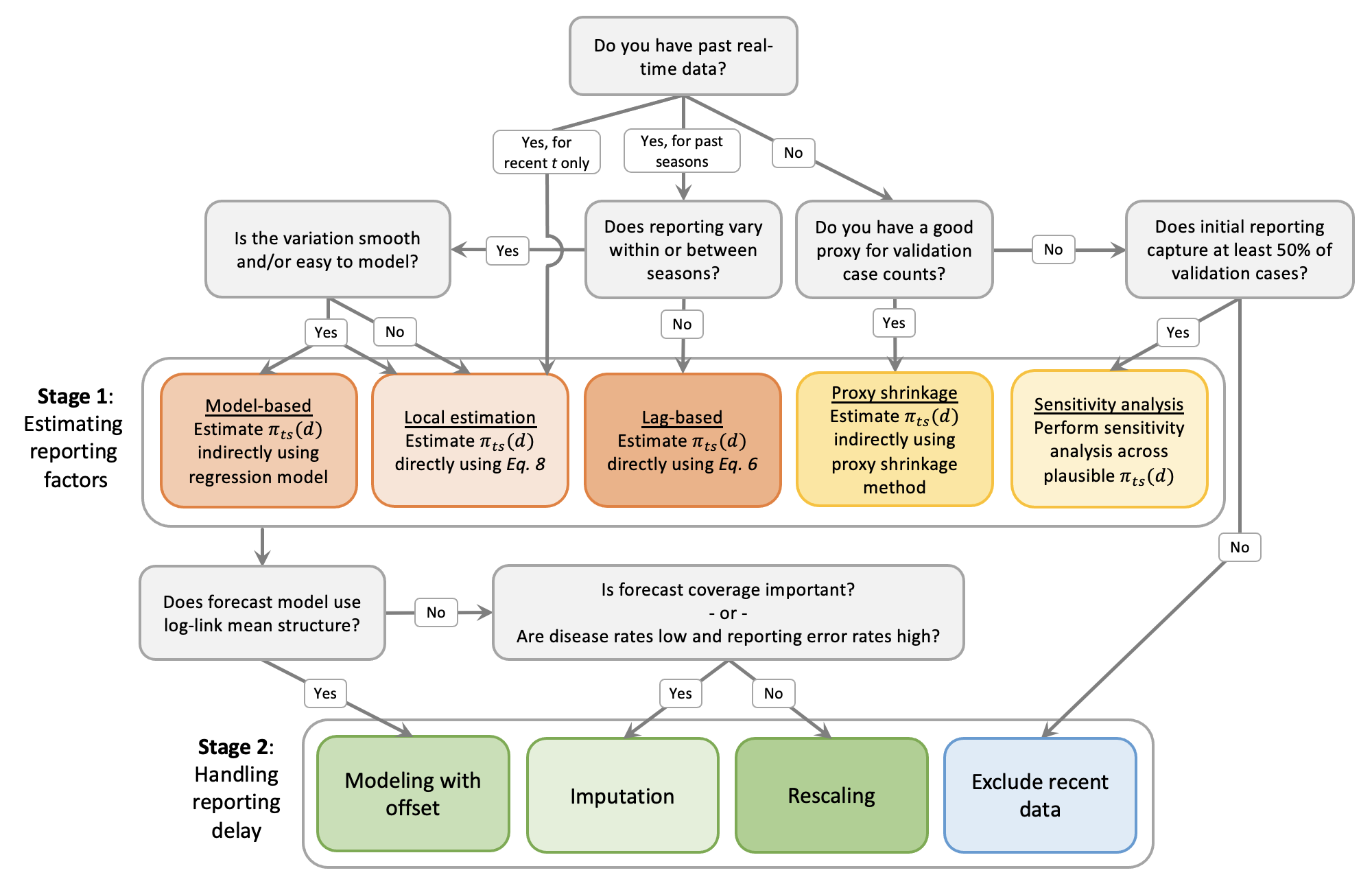}}
\label{methodsdiagram}
\end{figure}

In this section, we develop the notation and describe various methods for handling reporting delay in general. \textbf{Supp. Table B1} provides a summary of the notation. Suppose we consider time series data of reported cases for an infectious disease of interest, where cases are reported across multiple calendar time-points $t$ for each of multiple disease seasons $s$. For example, we may consider data on weekly case reports of seasonal influenza-like illness diagnoses, reported across several seasons. Throughout this paper, we will use ``week" to refer to the calendar time-point within each  season and ``year" to refer to the disease seasons, but these methods can be applied to more general time-scales. For a given calendar week $t$ in a given year/season $s$, the first reported case counts are often inaccurate, and the final ``validation" counts result from revisions at later dates. Let $d$ denote the number of weeks since the first reported case counts for a given calendar week. The number of weeks since the week's first report will also be called the ``lag" in reporting. Define $N_{ts}(d)$ to be the reported case counts for week $t$ in season $s$ reported with lag $d$, where $N_{ts}(0)$ corresponds to the very first report and $N_{ts}(\infty)$ corresponds to the final/validation case counts. \\
\indent In this paper, we provide and compare strategies for forecasting future \textit{validation} case counts using error-prone real-time reported case data, $N_{ts}(d)$. For example, we may want to obtain a forecast of validation cases at calendar time $t=5$ for season $s$ (denoted $N_{5s}(\infty)$) using the most recently-reported data for the previous 4 calendar weeks ($N_{1s}(3)$, $N_{2s}(2)$, $N_{3s}(1)$, and $N_{4s}(0))$ along with case data from past seasons. The impact of and correction for reporting delay will naturally depend on how accurate the real-time case data is. Let $\pi_{ts}(d)$ denote \textit{expected} proportion of validation cases for calendar week $t$ in season $s$ that are reported by lag $d$ weeks as follows
\begin{align} \label{pidef}
&\pi_{ts}(d) = E\left( \frac{\mathbf{N}_{ts}(d)}{\mathbf{N}_{ts}(\infty)}\right),
\end{align}
where $\mathbf{N}_{ts}(d)$ and $\mathbf{N}_{ts}(\infty)$ may be viewed as random variables with data realizations $N_{ts}(d)$ and $N_{ts}(\infty)$. A value of $\pi_{ts}(d)$ less than 1 indicates under-reporting and a value greater than 1 indicates over-reporting, relative to the validation counts. As $d \rightarrow \infty$, we expect that $\pi_{ts}(d) \rightarrow 1$ and $N_{ts}(d) \rightarrow N_{ts}(\infty)$. The \textit{reporting factor} for week $t$ and season $s$ at lag $d$ is defined as $\frac{1}{ \pi_{ts}(d)}$. We emphasize that this reporting factor is a function of the lag week $d$, and we will assume there is some threshold $\tau$ such that $\pi_{ts}(d) = 1$ for all $d > \tau$. We will not know the reporting factors in practice, and additional work will be needed to estimate the degree of reporting error and to account for reporting delay in the analysis. \\
\indent In the remainder of this section, we describe a general two-stage strategy for addressing reporting delay, including (stage 1) estimating reporting factors and (stage 2) forecast modeling, accounting for under- or over-reporting. First, we will describe several conceptual strategies for how to implement forecast modeling while addressing reporting delay given estimates of the reporting factors. Then, we will provide corresponding strategies for estimating reporting factors. These methods are summarized in \textbf{Figure \ref{methodsdiagram}}. 

\subsection{Conceptual strategies for handling reporting delay} \label{main_methods}
In this section, we provide several conceptual strategies for handling reporting delay. In implementing each of these methods, we assume that the forecasting model structure is fixed beforehand. Several of these methods require estimates of $\pi_{ts}(d)$. We discuss how to obtain these estimates in the next section. 

\subsubsection*{Rescaling of real-time data} 
One general strategy for correcting reporting delay is to use the most recently-reported case counts, $N_{ts}(d)$, and assumptions about the relationship between real-time and validation data (i.e., assumptions about $\pi_{ts}(d)$) to predict $N_{ts}(\infty)$ as follows:
\begin{align} \label{lagscaling}
&\hat N_{ts}(\infty) = \frac{N_{ts}(d)}{\pi_{ts}(d)},
\end{align}
where $\pi_{ts}(d)$ is replaced with an estimate \citep{Lawless1994}. Rescaling strategies are often used to correct for general case under- or over-reporting in disease modeling literature, but a key difference from simple rescaling approaches is that the correction factors are functions of the lag $d$ and possibly $t$, $s$, and additional covariates. Given a prediction $\hat N_{ts}(\infty)$ from \ref{lagscaling}, we can then fit our forecasting model using $\hat N_{ts}(\infty)$ in place of the real-time data.

\subsubsection*{Modeling real-time data with mean model offset}
The rescaling method involves pre-processing the real-time data to obtain predictions of the validation data for week $t$ in season $s$. Then, the predictions are input into the forecast modeling pipeline. An alternative strategy involves specifying a forecasting model for the real-time data directly, accounting for the reporting delay in the model structure \citep{England2002, McGough2020}. Usual implementations involve specifying a model for the \textit{incremental} cases reported for a given calendar week across lag times $d$, conditional on the cases reported up to lag time $d-1$ \citep{England2002}. These incremental case models are linked together into a chain that ultimately can be used to predict future values for $N_{ts}(\infty)$. Letting $n_{ts}(d) = N_{ts}(d) - N_{ts}(d-1)$ be the \textit{data realization} of the random incremental counts $\mathbf{n}_{ts}(d)$ reported on lag week $d$, we can model the random incremental counts using a negative binomial or Poisson regression model with the following mean structure:  
\begin{align} \label{meanmodel_chain}
&\text{log}(E(\mathbf{n}_{ts}(d))) = \log(E(\mathbf{N}_{ts}(\infty))) + \log(y_{ts}(d)),
\end{align}
where $y_{ts}(d) = \pi_{ts}(d) - \pi_{ts}(d-1)$ corresponds to the expected proportion of eventually-reported cases that are reported on lag week $d$. This model has the same mean structure as a corresponding model for the validation case counts, $\mathbf{N}_{ts}(\infty)$, but with an offset term. A key limitation of this formulation is that $y_{ts}(d)$ must be greater than zero; in other words, we can only apply this method when we have positive increments. In general, however, we may have over-reporting of cases in addition to under-reporting. Using logic in \textbf{Supp. Section C}, we instead propose modeling the cumulative counts directly using a negative binomial or Poisson regression model with mean structure as follows:
\begin{align} \label{meanmodel}
&\text{log}(E(\mathbf{N}_{ts}(d))) = \log(E(\mathbf{N}_{ts}(\infty))) + \log\left( \pi_{ts}(d)\right).
\end{align}
This mean structure formulation requires only that $\pi_{ts}(d) > 0$ for all $d > 0$. Individual increments, however, may increase or decrease. In implementing mean model offset approaches like \ref{meanmodel_chain}, researchers often estimate $\pi_{ts}(d)$ and the disease model parameters \textit{jointly} and specify a prior distribution for $y_{ts}(d)$ or $\pi_{ts}(d)$ \citep[e.g.,][]{McGough2020}. To make the estimation strategy more compatible with existing out-of-the-box software, however, we propose pre-estimating $\pi_{ts}(d)$ and then fitting the disease model treating $\log\left(\hat\pi_{ts}(d)\right)$ as a fixed offset in the mean model. Alternative distributional assumptions for $\mathbf{N}_{ts}(d)$ may also be considered (e.g., lognormal as in \citet{Kuang2020}), but this offset-based method is limited to settings where mean counts are modeled using some type of log link function. Compartmental susceptible-infectious-recovered models, for example, are not compatible with this method.

\subsubsection*{Imputation of validation case counts}
The rescaling method involves replacing the real-time data with a predicted value of $N_{ts}(\infty)$ prior to disease modeling. However, this approach does not capture the \textit{uncertainty} in this prediction. Recasting this approach in a missing data framework, the rescaling method predicts $N_{ts}(\infty)$ using the \textit{expected value} of $\mathbf{N}_{ts}(\infty)$ given the observed data. It is well-known in the missing data literature, however, that so-called conditional mean imputation may result in under-coverage of model parameters \citep{Little2002}. Similarly, we could see under-coverage of resulting disease model forecasts. \\
\indent To address this challenge, we can perform disease modeling based on \textit{draws} or \textit{imputations} of the validation data \citep[e.g.,][]{Hohle2014, Brooks2018}. In order to keep the handling of reporting delay separate from the disease modeling, we propose obtaining $M$ multiple imputations/draws of the validation data for the current season up to the current calendar time. Fixing these imputations, we can then fit the forecasting model to each of the $M$ imputed datasets and obtain $M$ forecasts (e.g., 1-week forecasts) and corresponding standard errors. We then combine results \textit{across} imputed datasets using Rubin's multiple imputation combining rules \citep{Little2002}. Additional details about this algorithm can be found in \textbf{Supp. Section D}.\\
\indent Many different approaches can be used to define a distribution for imputing validation case counts given the available data. Using results from the actuarial literature as in \textbf{Supp. Section C}, we propose imputing validation case counts from the following truncated normal distribution:   
\begin{align} \label{midistribution}
& \mathbf{N}_{ts}(\infty) \vert N_{ts}(d) \sim TruncNormal\left(\frac{N_{ts}(d)}{ \pi_{ts}(d)}, \frac{\vert 1-  \pi_{ts}(d)\vert}{  \pi_{ts}(d)^2}  N_{ts}(d) ; l, u\right),
\end{align} 
where $\pi_{ts}(d)$ is replaced with an estimate and where truncation limits $l$ and $u$ restrict imputed values to be greater than $N_{ts}(d)$ if $ \pi_{ts}(d) < 1$ and less than $N_{ts}(d)$ if $ \pi_{ts}(d) > 1$. A key property of this distribution is that the variance of the imputed validation data decreases as $\pi_{ts}(d)$ increases, meaning that imputed $N_{ts}(\infty)$ will be closer to $N_{ts}(d)$ when the expected amount of reporting error is small. The truncated normal distribution in \ref{midistribution} will have expectation different than $\frac{N_{ts}(d)}{ \pi_{ts}(d)}$ due to non-symmetric bounds. In \textbf{Supp. Figure D2}, we demonstrate that the percent error in this expectation relative to $\frac{N_{ts}(d)}{ \pi_{ts}(d)}$ is small, particularly in settings with a large amount of reporting error.   \\
\indent A primary limitation of the multiple imputation approach is that it requires the forecasting model to be fit multiple times. For many slow estimation methods, this may not be feasible. When the forecasting model involves Markov Chain Monte Carlo (MCMC) estimation (e.g., some Bayesian models), however, a simple alternative is to handle the missing data within the MCMC estimation algorithm, where a \textit{single} imputed value for each missing $N_{ts}(\infty)$ is generated using \ref{midistribution} within each MCMC iteration as in \citet{Hohle2014}. Parameters are then drawn within each iteration, conditioning on the imputed validation data. The resulting posterior forecast distributions can then be used directly. 

\subsubsection*{Exclusion of most recent data reports}
In some cases, the most recently reported real-time data may be too error-prone to contribute meaningfully to forecasting given the information available to the forecaster. In this setting, we propose excluding the most recently reported data from modeling and forecasting.

\subsection{Estimating reporting factors} \label{reporting}
For many of the methods described in \textbf{Section \ref{main_methods}}, 
the challenge of handling reporting delay boils down to specification of a relationship between the real-time and validation data, i.e., $\pi_{ts}(d)$. Below, we describe several different strategies for estimating $\pi_{ts}(d)$ using either real-time case reports for past seasons or by leveraging external data to directly nowcast $N_{ts}(\infty)$. 

\subsubsection*{Reporting factors as a function of lag only}
 A common assumption in the literature is that the reporting factors are constant in $t$ and $s$. Using data from past seasons/weeks in which \textit{both} real-time and validation data are available, we estimate the inverse reporting factors as 
\begin{align} \label{preest}
& \hat \pi_{ts}(d) = \hat \pi(d) = \frac{\sum_{i,j} N_{ij}(d)}{\sum_{i,j} N_{ij}(\infty)},
\end{align} 
with $\hat \pi(d) = 1$ for all $d$ greater than some fixed threshold, $\tau$. More sophisticated estimation strategies using parametric assumptions for $\pi(d)$ can also be used, as discussed in \citet{England2002}. An alternative strategy for estimating $\pi(d)$ would be to average $\frac{N_{ts}(d)}{N_{ts}(\infty)}$ across $s$ and $t$. Although not shown, simulations demonstrated that this proportion averaging approach tended to produced unstable $\hat \pi(d)$ with larger variability.

\subsubsection*{Reporting factors based on regression modeling of real-time data} 
The estimator in \ref{preest} assumes that the reporting factor depends on $d$ only, but we can imagine settings where the reporting may vary across seasons ($s$) and/or within seasons ($t$). Additionally, we may have covariates $X$ predictive of the difference between real-time and validation case reports (e.g., number of patients tested for the disease or disease-related Google search volumes). In this setting, we propose estimating $\pi_{ts}(d)$ indirectly by modeling the validation case counts as a function of available data, the lag $d$, the season $s$, the week $t$, and/or covariates $X$. For example, we may fit a Poisson regression model with a log link, adjusting for lag, season, week, and $X$ and using the log of the available counts as an offset in the mean model. See \textbf{Supp. Section F} for an example. Let function $f$ represent the estimated mapping between the real-time data and the validation data. We can estimate the inverse reporting factors as
\begin{align} \label{modelcorrection}
&\hat \pi_{ts}(d) =\frac{N_{ts}(d)}{f(N_{ts}(d); X)},
\end{align}
where $\hat \pi_{ts}(d) = 1$ for $d>\tau$ and where $\hat N_{ts}(\infty) = f(N_{ts}(d); X)$. 

\subsubsection*{Reporting factors using ``local" real-time data}
 The previously-described methods leverage past season data on the relationship between real-time and validation case counts to estimate reporting factors. However, these methods may perform poorly if reporting in the current season differs substantially from reporting in previous seasons in an unpredictable way (i.e., not easily predicted based on past-seasons' reporting trends). In this setting, we would like to use real-time data from the most recent weeks to estimate reporting factors that better capture current reporting practices. The challenge, however, is that validation data are not yet available for recent weeks. \\
 \indent We propose using the \textit{changes} in available case counts across $d$ to obtain a conservative estimate of $\pi_{ts}(d)$ as follows:
\begin{align} \label{local_preest}
& \hat \pi_{ts}(d) = \frac{\sum_{i = t - K}^{t-d-1} N_{is}(d)}{\sum_{i = t - K}^{t-d-1} N_{is}(t-i)},
\end{align}
  where the numerator contains past weeks' available data and the denominator contains the most recently-reported data for the previous $K$ weeks. Define $\tau$ to be the lag value after which we assume true $\pi_{ts}(d) = 1$ for all $d > \tau$. For this estimator, we define $\hat \pi_{ts}(d) = 1$ for $d>min(K,\tau)$. $K$ can take any positive integer value, but we recommend setting $K = \tau$ in practice.  \\
 \indent The estimator in \ref{local_preest} is conservative in that it is \textit{biased toward 1} relative to the true $\pi_{ts}(d)$ when reporting is such that (1) $N_{ts}(j) \leq N_{ts}(j+1)$ for $j \geq 0$ (monotone under-reporting) or (2) $N_{ts}(j) \geq N_{ts}(j+1)$ for $j \geq 0$ (monotone over-reporting). This bias comes from using the most recent case counts rather than the corresponding validation values in the denominator. Despite this bias, \ref{local_preest} may provide a more accurate estimate of $\pi_{ts}(d)$ than the previously-described approaches when the current weeks' reporting is very different from reporting in past seasons. 

\subsubsection*{Reporting factors based on proxy shrinkage} 
The preceding methods assume some prior information is available about the relationship between the real-time and validation case counts. For many diseases however, historical data on validation case counts are readily available but the real-time reports for past seasons are either unavailable or incomplete. In this setting, we propose leveraging external data sources to estimate $\pi_{ts}(d)$. \\
\indent Suppose we have historical data on the relationship between validation case counts $N_{ts}(\infty)$ and some external data, collectively denoted $p_{ts}$, that are associated with $N_{ts}(\infty)$. Examples of external data include Google search volumes, social media mentions (e.g., from Twitter), and Wikipedia searches \citep{Dugas2013,Hickmann2015}.Using these historical data, we fit a nowcast model for validation case counts given $p_{ts}$. This model could be complicated and include data from many different sources, incorporate penalization, etc. Let $g$ denote the mapping between $p_{ts}$ and model-predicted validation case counts. These predictions may be viewed as an error-prone \textit{proxy} for the current season validation cases. \\
\indent Using this external data model proxy, we propose estimating the validation case counts as a weighted average of the proxy and the observed real-time data as follows
\begin{align} \label{proxyexpression}
&\hat N_{ts}(\infty) = w_d g(p_{ts}) + (1-w_d) N_{ts}(d),
\end{align}
and we estimate corresponding inverse reporting factors as $\hat \pi_{ts}(d) = \frac{N_{ts}(d)}{\hat N_{ts}(\infty)}$. For large $d$, we expect $N_{ts}(d)$ and $N_{ts}(\infty)$ to be very close. Therefore, we will define weights $w_d$ such that $w_d \rightarrow 0$ as $d \rightarrow \infty$ and impose $w_d  = 0$ for all $d > \tau$. As the number of lag weeks increases, the shrinkage estimator will put more and more weight on the observed data. In a related method, \citet{Osthus2019b} proposed defining $w_d = \left(\frac{\omega}{d + 1}\right)^2$ with $\omega = 0.75$. The choice of weight and performance of this method will depend on the degree of reporting delay and how well the proxy model $g$ predicts the validation case counts.

\subsubsection*{Sensitivity analysis}
All of the above methods for estimating reporting factors require either historical real-time data or a good proxy for the validation case counts based on external data. In practice, we may not have either available. In this case, one approach for implementing the methods in \textbf{Section \ref{main_methods}} is to perform multiple parallel analyses assuming different plausible reporting mechanisms in a sensitivity analysis framework. This approach can be used to evaluate the robustness of forecasts to assumptions about the delay mechanism.

\section{Application to Dengue Fever and Seasonal Influenza Forecasting }  \label{apppres}
\indent We considered data on dengue fever cases in Puerto Rico (1990-2009) and for seasonal influenza-like illness (ILI) cases in the United States (2010-2011 through 2018-2019 seasons). Our goal was to evaluate how the quality of disease forecasts is impacted by the handling of reporting delay. With the exception of the mean model offset method, all of the methods in \textbf{Figure \ref{methodsdiagram}} for handling reporting delay can be applied in conjunction with any forecasting model. For demonstration, we considered two reasonable forecasting model settings: an auto-regressive moving average (ARMA) model for log-cases \citep[e.g.,][]{Kandula2019} and a Bayesian Gaussian process-based model (hereafter, referred to as Inferno from \citet{Osthus2021}). Both models assumed a log link mean structure. Additional details about these forecast models are provided in \textbf{Supp. Section E}. 

\subsection{Obtaining weekly forecasts, accounting for reporting delay} \label{perf}
\indent For each week after the first two seasons, we constructed an artificial dataset consisting of the real-time and historical observations that would have been available to use for forecasting. First assuming that reporting does not vary over $t$ and $s$, we estimated reporting factors $\hat \pi(d)$ from \ref{preest} using the previous 2 years' data. To allow estimated reporting factors to vary by $t$ and $s$, we also estimated $\hat \pi_{ts}(d)$ indirectly as in \ref{modelcorrection} by fitting a Poisson regression model with a log link to the prior observed validation data, including $d$, $s$, and a 3-degree natural spline of $t$ as predictors and treating the available data $\log(N_{ts}(d))$ as an offset. Additional details and diagnostics can be found in \textbf{Supp. Section F.1}. Finally, we estimate ``local" $\pi_{ts}(d)$ using \ref{local_preest} and the previous $K$ weeks' real-time reported data (using $K = 15$ for ILI and $K=6$ for dengue fever). As an additional sensitivity analysis, we also estimate ``local" $\pi_{ts}(d)$ across a range of $K$ values between 2 and 100 for each dataset.  \\
\indent Fixing these reporting factors and using either an ARMA(2,2) or Inferno forecasting model structure, we then fit the forecasting model using each of the rescaling, model offset, and imputation methods in \textbf{Section \ref{main_methods}}. For ARMA, the imputation method was implemented using 10 imputed datasets and \ref{midistribution}. For Inferno, imputation using \ref{midistribution} was implemented within a Bayesian MCMC estimation algorithm. We also fit the forecasting model to the observed real-time data without correction and after excluding the most recent 1-3 weeks of data. To benchmark the performance of these methods, we also fit the forecasting model using the validation data. Inferno model estimation was based on 2500 MCMC iterations after a burn-in of 2500 iterations. \\
\indent For each forecast fit, we obtained a prediction for the validation case counts for the current week (nowcast) and for 1 and 4 weeks ahead (forecasts). We also calculated corresponding 50\%, 67\%, 95\%, and 99\% prediction intervals. We repeated this process for each calendar week of available data. We then summarized the predictive performance using the metrics described in \textbf{Section \ref{perf}}, aggregating across all calendar weeks and seasons for each dataset and forecast model structure separately. Results are also provided for an ensemble based on an equal weight linear combination of forecasts from the 13 methods (excluding validation data analysis). For the Inferno model, ensemble forecasts and corresponding prediction intervals were obtained using the stacked draws across methods. For the ARMA model, we obtained 2500 draws of predicted log-case counts using the Gaussian prediction distribution for each of the methods. Ensemble forecasts and prediction intervals for the validation case counts were then obtained using these stacked draws.

\subsection{Performance Metrics} \label{perf}
We evaluated the performance of the various methods based on properties of nowcasts (i.e., predicted validation values for the current week) and 1 and 4 week forecasts (i.e., predicted validation case counts for future weeks). Absolute prediction error for a given nowcast/forecast was defined as the absolute difference between the model estimate and the validation case counts. For assessing the accuracy of the point and interval estimates jointly, we calculated the weighted interval score from \citet{Bracher2021} as follows:
\begin{align} \label{wiseq}
&WIS(y) = \frac{1}{K + 0.5} \left[ 0.5 \vert y - \hat f_{ts}\vert +\sum_{j=1}^K \frac{\rho_j}{2}IS(j, y)\right]\\
& IS(j, y) = (u_j - l_j) + \frac{2}{\rho_j}(l_j - y)I(y < l_j) + \frac{2}{\rho_j}(y-u_j)I(y > u_j),
\end{align}
where $K$ is the number of intervals (in our case, 4), $\rho = (0.5, 0.33, 0.05, 0.01)$, $u_j$ and $l_j$ are the upper and lower confidence/credible interval limits corresponding to level $1-\rho_j$, and $\hat f_{ts}$ is the median of the nowcast/forecast distribution. We evaluated this the validation values. We also calculated the empirical coverage of the 95\% prediction intervals using the forecasts and validation case counts across all calendar weeks.

\subsection{Results} \label{real_results}
\begin{figure}[h!]
  \centering
\caption{Performance of nowcasts and forecasts in the Puerto Rico dengue fever and national US influenza-like illness data across all weeks $^1$}
\subfloat[Average absolute bias in nowcasts and 1 and 4 week forecasts]{\includegraphics[trim={0cm 0.5cm 0cm 0cm}, clip, height=2.2in]{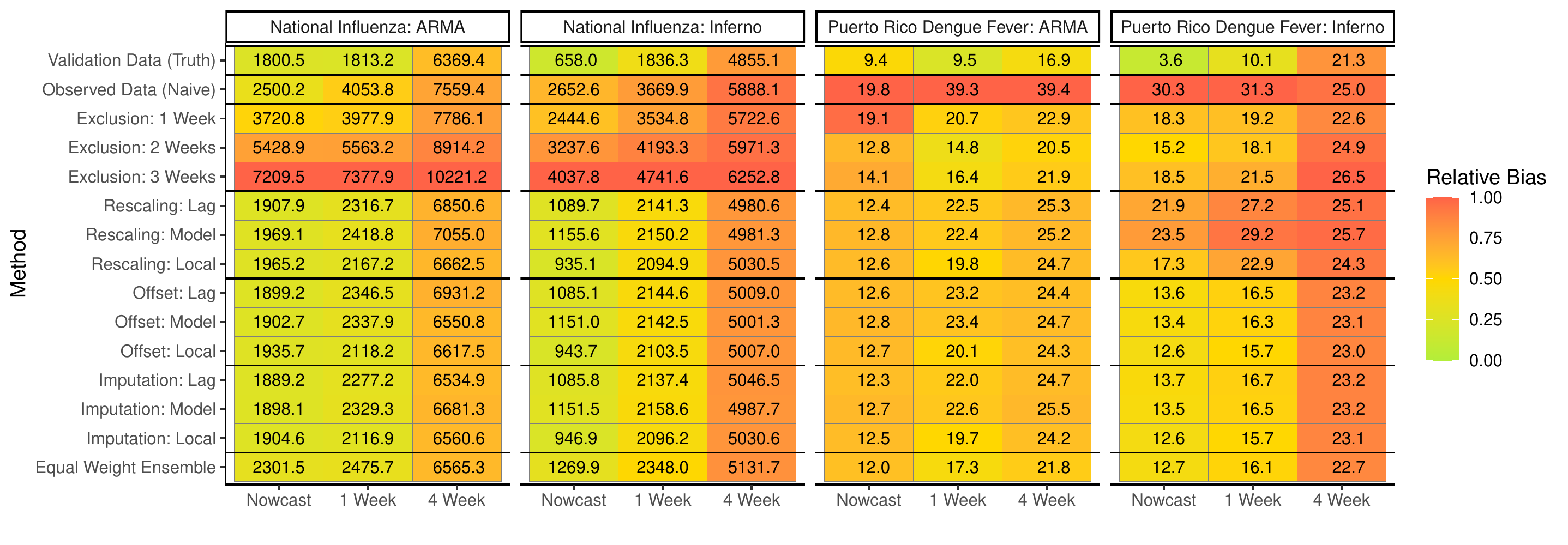}}\\
\subfloat[Coverage of 95\% prediction intervals for 1 and 4 week forecasts]{\includegraphics[trim={0cm 0.5cm 0cm 0cm}, clip, height=2.15in]{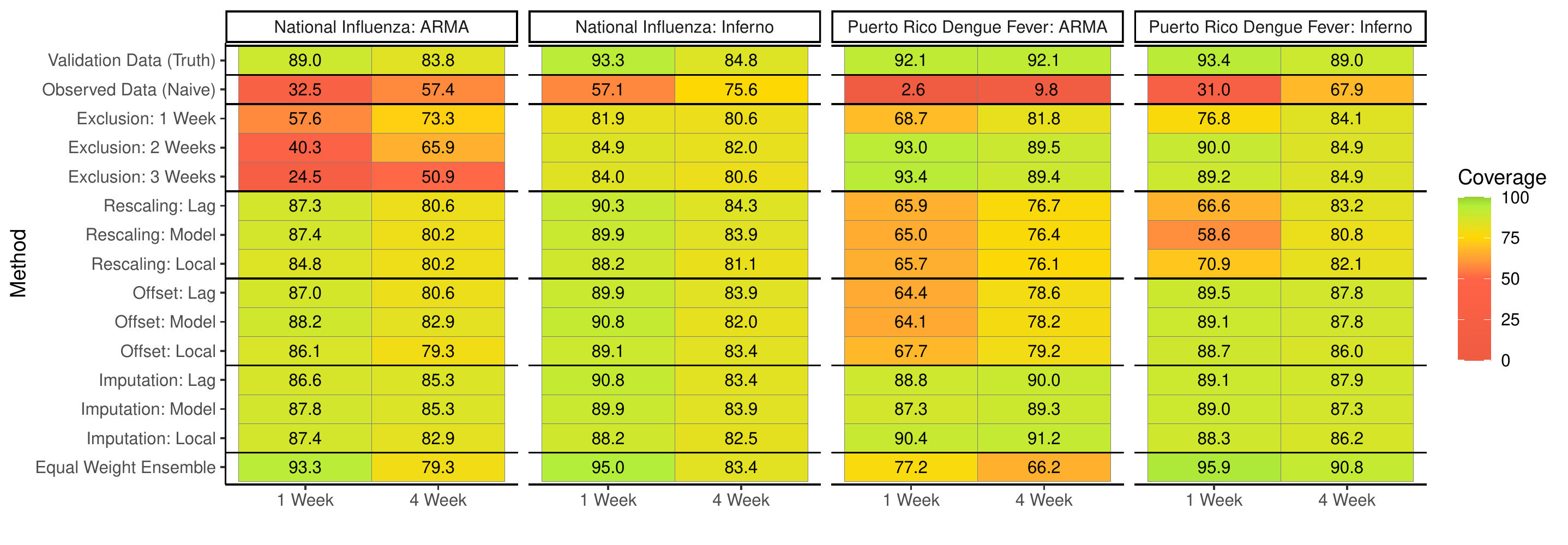}}
\caption*{\footnotesize $^1$ Results for dengue fever are aggregated across each of 50 weeks in 18 seasons (1992-2009). Results for US national influenza are aggregated across 35 weeks in 7 seasons. The ensemble method corresponds to an equal-weight linear combination of all methods except validation data analysis. ``Model" indicates that reporting factors were estimated via regression and allowed to vary by $t$ and $s$. ``Lag" indicators that reporting factors were estimated via \ref{preest}. ``Local" indicators that reporting factors were estimated via \ref{local_preest}. Relative absolute biases are calculated relative to the largest value in each column.}
\label{analysisaggregate_1}
\end{figure}

\begin{figure}[h!]
  \centering
\caption{Proportion of weeks in which each of 4 methods performs best in terms of 1-week forecast weighted interval scores}
\includegraphics[trim={0cm 0cm 0cm 0.7cm}, clip, width=6in]{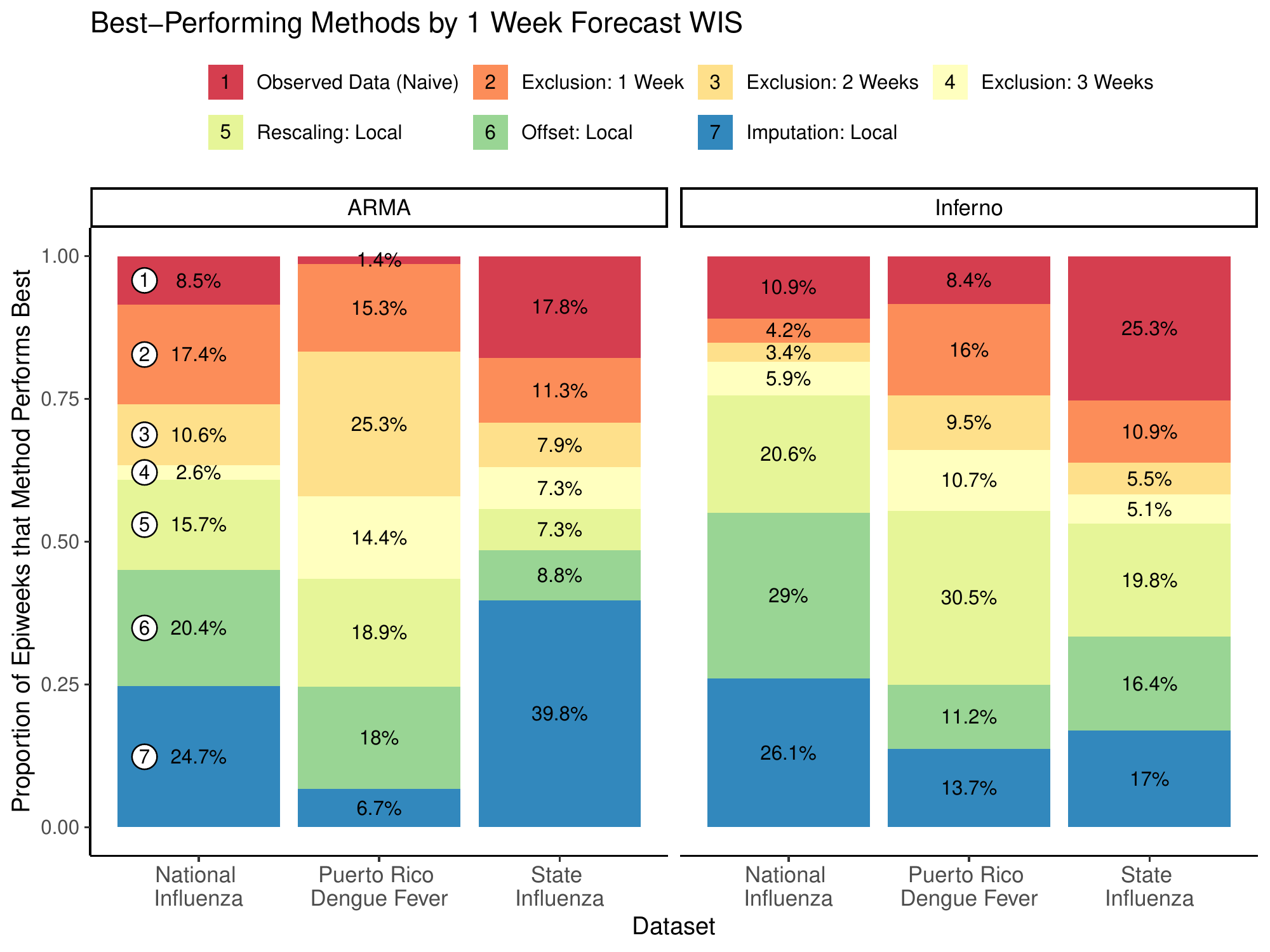}
\caption*{ \footnotesize  $^1$ Results for Dengue fever in Puerto Rico are aggregated across all 50 weeks in 18 calendar years (1992-2009). Results for US national influenza are aggregated across 35 weeks in 7 seasons (2012-2013 through 2018-2019). Results for US state-level influenza are aggregated across 35 weeks in 49 states (excluding Florida) for the 2018-2019 season.}
\label{analysisrankings}
\end{figure}

\underline{Prediction error and coverage:} \textbf{Figure \ref{analysisaggregate_1}} presents the average absolute nowcast/forecast errors for the dengue fever and US national ILI outcomes along with corresponding coverages of 95\% prediction intervals. Weighted interval scores are presented in \textbf{Supp. Figure F2}. For both datasets and forecast model structures, analysis of the observed data without correction resulted in large undercoverage (2-60\%) and higher prediction error relative to analysis of the validation data. Compared to uncorrected analysis of the observed data, the exclusion method (1-3 weeks) resulted in improved coverage and similar or reduced prediction error across all endpoints for the dengue fever data. In contrast, the exclusion method resulted in similar or increased prediction error in the national US ILI data. \\
\indent The rescaling, offset, and imputation methods uniformly resulted in higher coverage and similar or reduced prediction error relative to uncorrected observed data analysis for the dengue fever and US national ILI data. For the dengue fever data, the imputation methods tended to give better coverage than the rescaling methods (e.g., around 65\% vs nearly 90\%). In general, the rescaling method can be unstable (i.e., produce high error) in the dengue fever data, particularly for the Inferno forecast model (\textbf{Supp. Figure F3}). As described in \textbf{Supp. Section G}, this is due to a combination of very high under-reporting and a low disease rate, which results in rescaled validation estimates that are very sensitive to small fluctuations in the observed case counts. Both the mean model offset method and the imputation method resulted in stabler nowcasts/forecasts in this setting. \\
\indent The equal weight ensemble method attempts to produce more stable forecasts than any individual method by borrowing information across many methods. For both the Inferno dengue fever model and both ARMA and Inferno national influenza models, the ensemble estimation resulted in similar or better coverage than any of the individual methods, particularly for nowcasts. Resulting nowcast and forecast errors were generally between errors from the exclusion and rescaling/offset/imputation methods for the ILI data and similar or smaller than errors for other methods for the dengue fever data. Across both datasets, there is little evidence of improved forecast/nowcast performance for model-based vs. lag-based estimation of reporting factors. However, we observed lower forecast biases when reporting factors were estimated using \textit{local} reporting data. We did not formally assess the performance of proxy-based reporting factor estimation for these data.    \\ 
\indent In \textbf{Supp. Figures F4 and F5}, we present the relative accuracy of several methods across individual weeks during follow-up. We find that the relative performance of uncorrected data analysis tended to improve for the dengue fever data when validation case counts were very small, suggesting that the correction methods may over-correct in this setting. Similarly, uncorrected analysis of the US national ILI data gave improved relative performance shortly after the season peak, suggesting that the correction methods may be less able to adapt to sharp decreases in validation case counts. \\
\\
\underline{Comparative rankings:} We then calculated the proportion of weeks across seasons in which each of 7 methods performed best in terms of weighted interval scores for 1 week forecasts. For this analysis, we focused on local estimation of $\pi_{ts}(d)$ as it generally resulted in similar or better performance than the same methods using lag-based or model-based estimation. \\
\indent If all methods performed equally, we would expect each method to be the best performer about 14\% of the time (95\% confidence interval of [12\%,16\%] for dengue fever and state-level ILI and [10\%,18\%] for national ILI, assuming independent trials). As shown in \textbf{Figure \ref{analysisrankings}}, however, we saw that corrected analysis tended to outperform uncorrected analysis of the observed data for dengue fever and US national ILI data. For these data sources, these results suggest that we are almost always better off performing some kind of correction strategy to handle the reporting delay. Among the correction methods, the imputation and offset methods tended to have similar or better performance than the other methods in terms of 1-week forecast interval score rankings for the national influenza-like illness datasets. In contrast, the exclusion and rescaling methods had the best performance in the dengue fever data. \\
\indent Uncorrected analysis results for state-level ILI tell a different story than seen for national US ILI and dengue fever, where analysis of the observed state ILI data without correction had the best weighted interval score performance at least 17\%+ of the time. As shown in \textbf{Supp. Figures F6 and F7}, the relative performance of the delay correction methods varied between states, but the imputation-based correction method tended to have the best performance on average. For the Inferno model, uncorrected analysis outperformed all correction methods on average (lowest forecast weighted interval score 25\% of the time). \\
\indent Unlike in the dengue fever and national ILI data, the reporting delay mechanism sometimes changed \textit{substantially} between seasons for individual states (\textbf{Figures A1 and A2}). Although not shown, this resulted in reduced forecast performance for the strategies that relied on past season reporting data to correct delay in the current season. As shown in \textbf{Supp. Figure F8}, the local $\pi_{ts}(d)$ estimation method produced much better estimates of the inverse reporting factors for the current season. This serves as a cautionary note against using lag-based and model-based $\pi_{ts}(d)$ estimation methods when past season information on reporting practices is poorly representative of the current season.   \\
\indent For all datasets, we observed better comparative performance for naive observed data analysis with the Inferno model than with the ARMA model. This was due to (1) a tendency for the observed data Inferno model to perform particularly well relative to the other methods in weeks where the validation case counts were very low (\textbf{Supp. Figures F4, F5, and F9}) 
and (2) poor ARMA forecast performance across the board when analysis was based on real-time observed data (\textbf{Supp. Figure F10}). These results suggest that ARMA forecast models may tend to perform poorly when applied to real-time data subject to a large amount of reporting error. \\
\\
\underline{Weeks $K$ used for local $\pi_{ts}(d)$ estimation:} In the analyses presented in \textbf{Figure \ref{analysisaggregate_1}}, we apply the local estimation strategy in \ref{local_preest} with $K$ set equal to $\tau$, the number of weeks beyond which we assume $\pi_{ts}(d) = 1$. However, any positive integer value for $K$ could theoretically have been chosen. In \textbf{Supp. Figures F11-F13}, we present the $\pi_{ts}(0)$ estimates and corresponding nowcast/forecast performance obtained as a function of $K$ for the dengue fever, national US ILI, and Vermont ILI datasets. We demonstrate that the estimate in \ref{local_preest} looks more and more like the simple lag-based estimate using \ref{preest} as $K$ becomes large. The best nowcast and forecast performance was generally seen for $K$ near $\tau$, with larger $K$ resulting in particularly large increases in prediction errors for the Vermont data.

\FloatBarrier

\newpage

\section{Performance of methods for simulated real-time data}  \label{sims}
The analyses in \textbf{Section \ref{apppres}} provide intuition on the performance of the various reporting delay correction methods in three real data settings. A key limitation of several of the methods explored in \textbf{Section \ref{apppres}} is that they require high-quality data on real-time case reports from past seasons along with corresponding assumptions about how reporting varies across and within seasons. Since reporting delay for the current season is not well-understood at the time of forecasting in practice, it is important to evaluate the impact of these assumptions and violations on method performance. To further develop our intuition for when these methods will and will not perform well, we conducted a simulation study.  

\subsection{Simulation set-up} \label{sims_setup}
\indent We simulated validation disease counts to mimic observed disease rates of dengue fever in Puerto Rico between 1990 and 2009 as follows:
\begin{enumerate}
\item Let $\beta_{ts}$ be the three-week moving average of the validation case counts in the actual dengue fever data, and let $ \tau_t$ be the average of $N_{ts}(\infty)/ \beta_{ts}$ across all seasons. Define $\theta_{ts} = \beta_{ts}*\tau_{t}$ to be the mean disease rate in the simulated data.
\item We generated 200 simulated versions of the validation data for each $t$ and $s$ by drawing from a negative binomial$(r,p)$ distribution with $p = \frac{r}{r+\theta_{ts}}$ and $r = 100$. The hyperparameter $r$ was chosen to mimic the variability in actual dengue fever validation data. 
\end{enumerate}
\textbf{Supp. Figure H1a} provides a visualization of the resulting case counts across these 200 simulated datasets. For each simulated dataset, we then generated real-time data for each calendar week under different assumed reporting mechanisms. All reporting mechanisms followed the reporting profiles in \textbf{Supp. Figure H1b}, which were parameterized by constant $a$ between 0.05 and 1 representing the proportion of eventually reported events reported at lag 0. We considered the following simulation scenarios:
\begin{enumerate}
\item \textit{Constant}: reporting was constant in $t$ and $s$ and corresponded to $a = 0.05$.  
\item \textit{Vary by week}: reporting varied by $t$ and was constant in $s$. $a$ initially increased from 0.05 to 1 during weeks 1 to 25 in each season and then decreased from 1 to 0.05 thereafter.
\item \textit{Large improvement between seasons}: reporting substantially improved in the last season, with $a=0.05$ for 1990-2008 and $a=0.50$ for 2009.
\item \textit{Large worsening between seasons}: reporting substantially worsened in the last season, with $a=0.50$ for 1990-2008 and $a=0.05$ for 2009. 
\item \textit{All season combinations}: Each strata of 10 simulation replicates was assigned a different value for $a$ between 0.05 and 1. Within each strata, reporting (i.e., $a$) was constant in $s$ and $t$. 
\end{enumerate}
For each set of simulated validation data, we also simulated 4 external proxy variables such that $p_{ts} = 2\log(N_{ts}(\infty)+0.1) + e_{ts}$, where $e_{ts} \sim N(0,\sigma^2)$ and where $\sigma^2$ took values in (0.01, 1, 4, 16). These error rates corresponded to correlations between transformed $p_{ts}$ and $N_{ts}(\infty)$ of 0.99, 0.80, 0.50, and 0.13.

\subsection{Performance under Scenarios 1-4 in the simulated 2009 season} \label{sims1}
\begin{figure}[h!]
  \centering
\caption{Performance of proposed methods for handling reporting delay in 2009 across 100 simulated datasets using ARMA models$^1$}
\subfloat[Average absolute prediction error in nowcasts and forecasts]{\includegraphics[trim={0cm 0cm 0cm 0cm}, clip, height=3in]{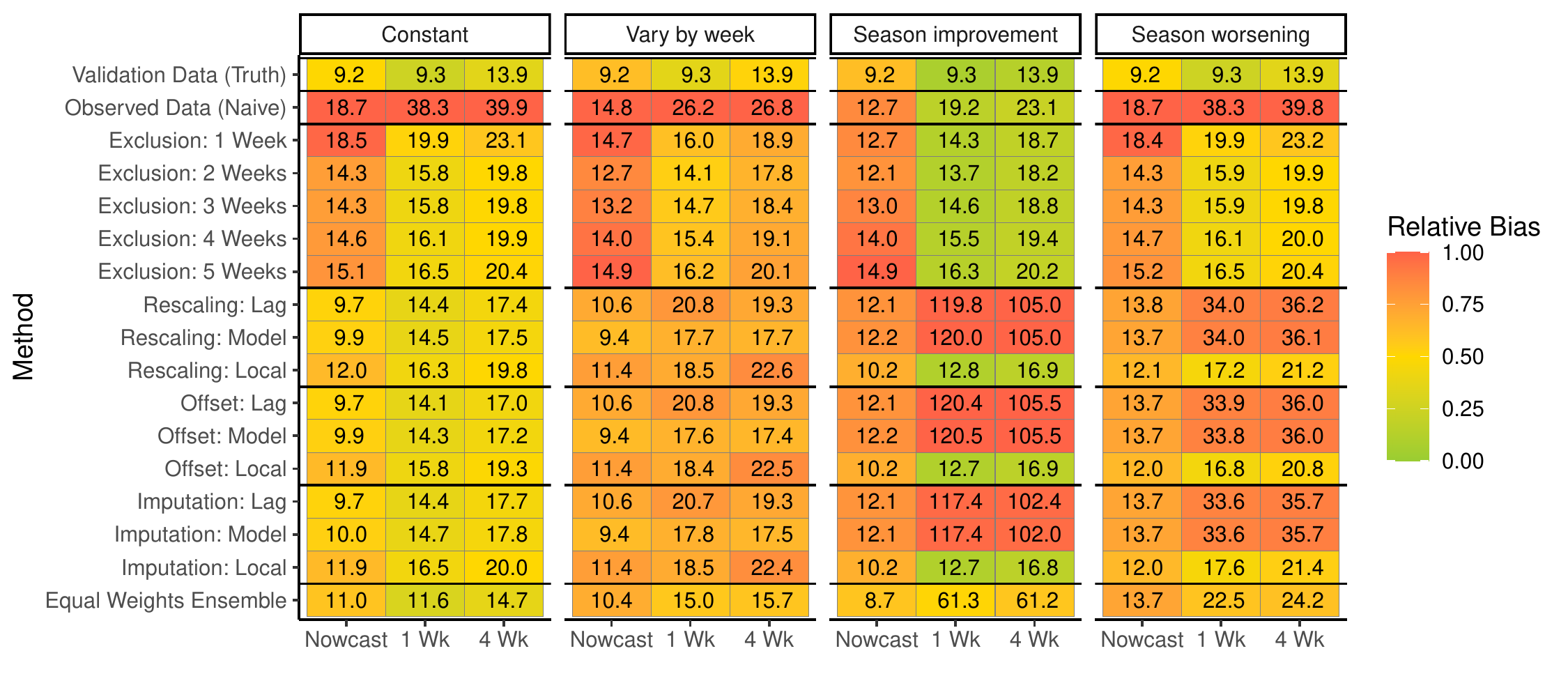}}\\
\subfloat[Coverage of 95\% prediction intervals for forecasts]{\includegraphics[trim={0cm 0cm 0cm 0cm}, clip, height=2.9in]{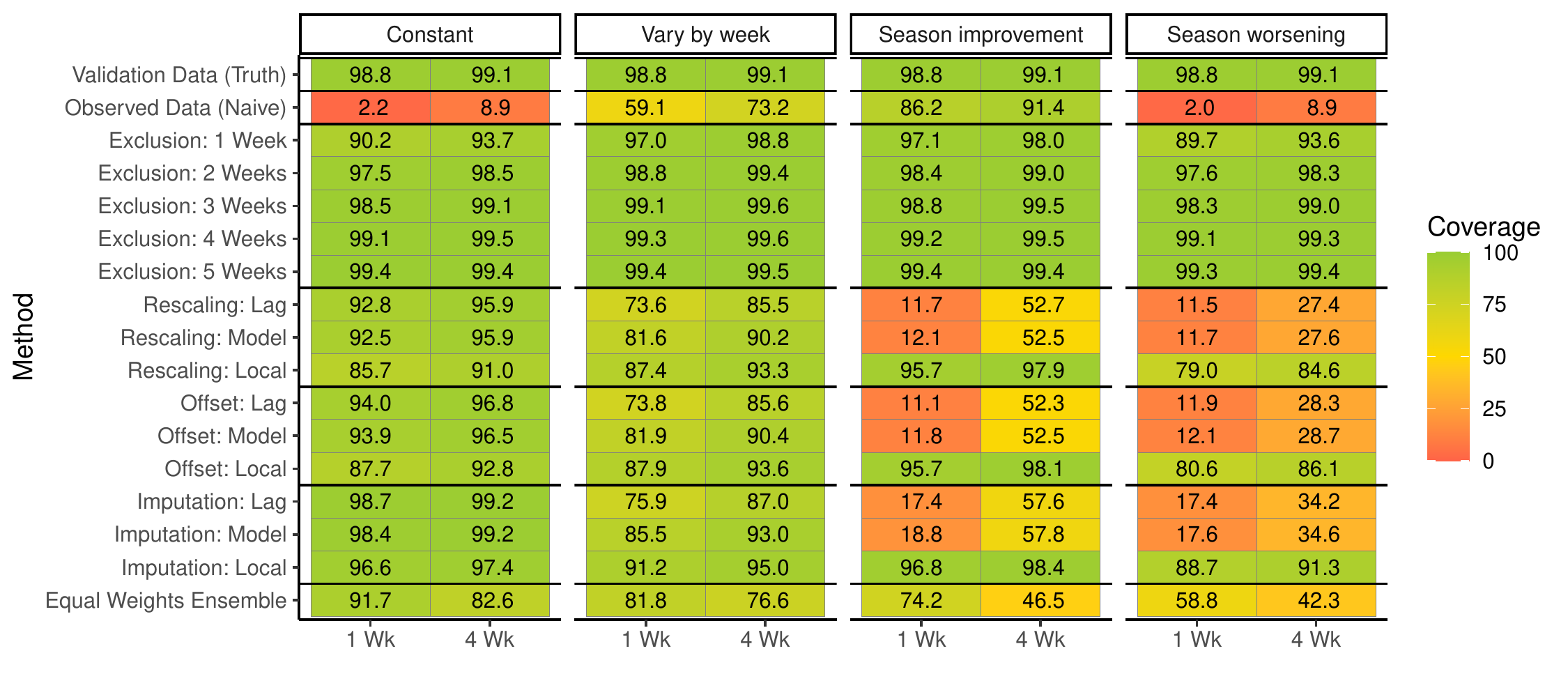}}
\caption*{\footnotesize $^1$ Results are aggregated across all 50 weeks in 100 replicate seasons. Each result, therefore, represents aggregates 5000 nowcasts or forecasts. When reporting factors varied across seasons, $\pi_{2007} = \pi_{2008} = (0.01, 0.05, 0.55, 0.85, 0.95, 0.98, 1)$ and $\pi_{2009} = (0.04, 0.54, 0.84, 0.0.94, 0.97, 0.99, 1)$. \\
\indent The ensemble method corresponds to an equal-weight linear combination of all methods except validation data analysis and exclusions of 4 and 5 weeks' data. ``Model" indicates that reporting factors were estimated via regression and allowed to vary by $t$ and $s$. ``Lag" indicators that reporting factors were estimated via \ref{preest}. ``Local" indicators that reporting factors were estimated via \ref{local_preest}. Relative absolute biases are calculated relative to the largest value in each column. Model-based $\pi_{ts}(d)$ estimation assumes reporting factors vary across weeks but incorrectly models \text{how} reporting factors vary across weeks. }
\label{sim_heatmaps}
\end{figure}

 In this simulation study, our goal was to compare the performance of each of the reporting delay correction methods when the reporting factor estimates were correctly specified (Scenario 1 and model-based $\pi_{ts}(d)$ for Scenario 2) and when they were mis-specified (lag-based $\pi_{ts}(d)$ for Scenario 2 and Scenarios 3-4). For this exploration, we focused our attention on the first 100 simulated datasets corresponding to 2009. \\
 \indent For each calendar week in 2009 and simulated datasets under Scenarios 1-4, we applied the procedure described in \textbf{Section \ref{perf}} to construct an artificial real-time dataset that would have been available for forecasting. For each artificial dataset, we applied the methods in \textbf{Figure \ref{methodsdiagram}} exactly as we did in the actual dengue fever data, but this time we repeated this analysis across 100 simulated versions of the data. We obtained nowcasts, forecasts, and corresponding 95\% prediction intervals, and we summarized these nowcasts/forecasts using the performance metrics in \textbf{Section \ref{perf}}. We then aggregated these metrics across the 100 simulated datasets and 50 calendar weeks in 2009 for each of the simulation scenarios. \\
\indent \textbf{Figure \ref{sim_heatmaps}} presents the average absolute nowcast and forecast prediction errors and corresponding coverage of 95\% prediction intervals obtained from applying the methods with a ARMA(2,2) forecast model. Weighted interval scores are shown in \textbf{Supp. Figure H2}. We summarize these results for each of the 4 simulation scenarios below. \\
\\
\noindent \underline{Reporting factors constant (Scenario 1):} All reporting correction strategies resulted in lower prediction error and higher coverage relative to analysis of the observed data without correction (e.g., coverages $<$ 10\% for uncorrected analysis compared to coverages over 90\%). We also see improvements over observed data analysis in terms of weighted interval scores for all reporting correction methods. The exclusion methods produced higher nowcast and forecast errors on average relative to the other correction methods. This indicates that the most recently reported cases can contribute useful information for nowcasting/forecasting even in the presence of extreme under-reporting as long as the reporting factors are correctly specified. Imputation methods resulted in higher nowcast and forecast coverage than the rescaling and model offset methods, although this coverage was overly conservative (greater than 95). Model-based estimation of $\pi_{ts}(d)$ resulted in a small increase in prediction error relative to lag-based estimation. More flexible estimation of $\pi_{ts}(d)$, therefore, may come with some small price in terms of prediction error when reporting is truly constant across $t$ and $s$. Local estimation of $\pi_{ts}(d)$ resulted in small additional increases in prediction error and weighted interval scores.  \\
\\
\noindent \underline{Reporting factors vary by week (Scenario 2):} When reporting factors truly varied within each season, correction methods using model-based estimates of $\pi_{ts}(d)$ outperformed use of lag-based estimates, which incorrectly assumed reporting did not vary over $s$ or $t$. However, the difference in performance was modest, even in this setting with extreme intra-season variation in reporting. The local $\pi_{ts}(d)$ estimation method performed well in this setting, with higher coverages than the model-based method but also with slightly higher prediction error and weighted interval scores. As before, the exclusion method resulted in higher nowcast error than the other correction methods, but the exclusion method did provide better coverages than the rescaling, offset, and imputation methods. \\
\\
\noindent \underline{Large reporting factor change between seasons  (Scenarios 3-4):} When reporting for the current season was very different than the previous seasons, reporting factor estimation using past season data (lag-based and model-based) resulted in poor estimates of the current season under-reporting. When the current season under-reporting was much \text{better} than previous seasons, use of \textit{past-season} reporting factors resulted in over-correction and over-estimation of the validation case counts, and we saw huge prediction errors and low coverage (e.g., 10-60\%) resulting from very poorly specified estimates of $\pi_{ts}(d)$ for the rescaling, offset, and imputation methods. When the current season under-reporting was much \textit{worse} than previous seasons, use of past-season reporting factors resulted in under-correction and under-estimation of validation case counts, but resulting prediction errors were still less than those from analysis of uncorrected observed data. Combined, these results indicate that we may be better off under-estimating the degree of under-reporting than severely over-estimating it.\\
\indent In contrast to the model-based and lag-based methods for estimating $\pi_{ts}(d)$, the local estimation method performed well with much higher coverages (e.g., $<20\%$ vs 95\% for 1 week forecasts) and much lower prediction errors for both Scenarios 3 and 4. The exclusion methods also had good performance. To summarize, the best performing methods were those that did not rely on past season data for handling of under-reporting.

 \clearpage
\subsection{Performance as a function of the reporting rate (Scenario 5)} \label{sims1}
\begin{figure}[h!]
  \centering
\caption{Coverage of 95\% prediction intervals for 1 week forecasts across various \textit{assumed} reporting profiles in 2008 and 2009 simulated datasets $^1$ }
\includegraphics[trim={0cm 0cm 0cm 0.8cm}, clip, width=6in]{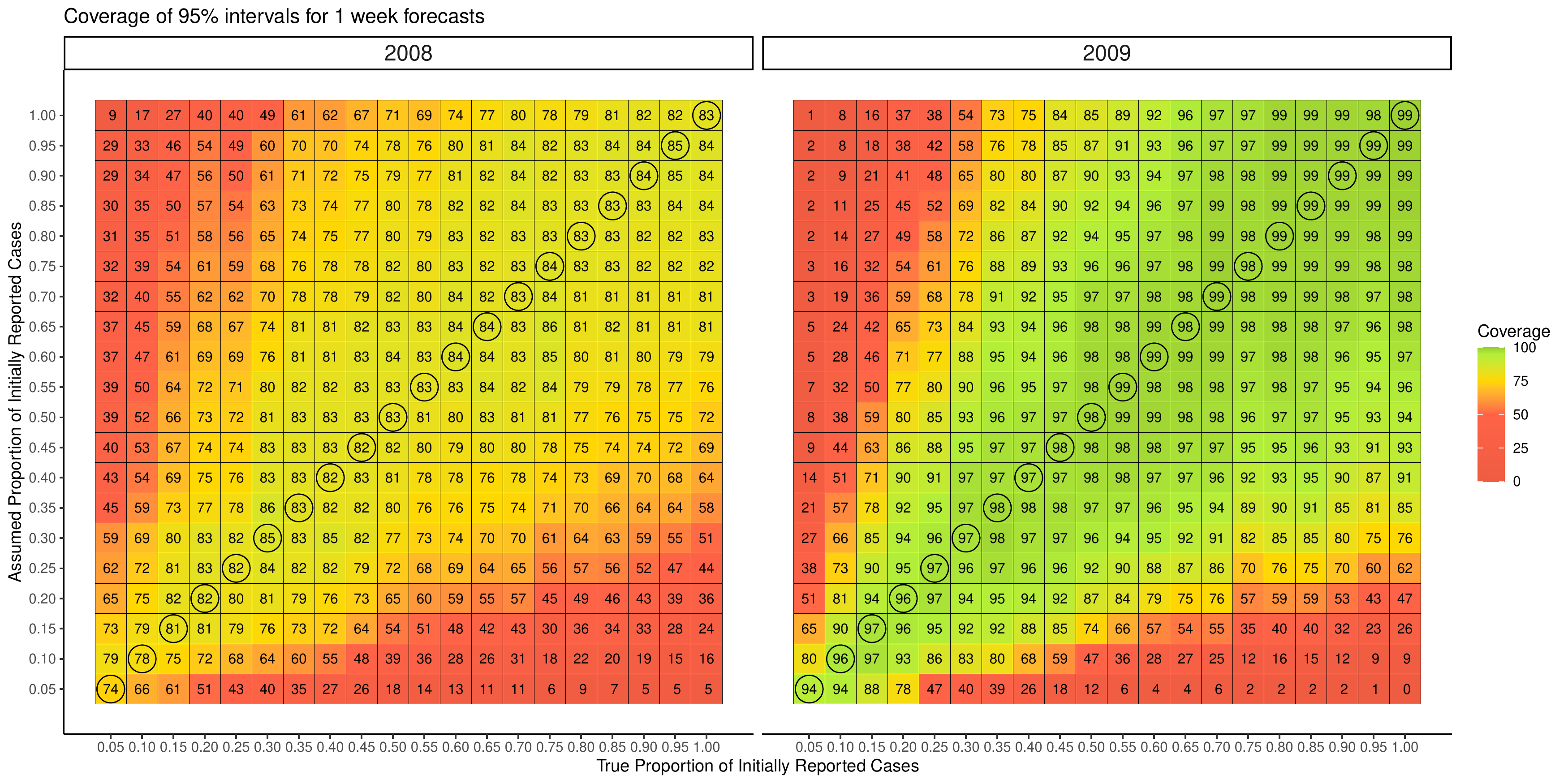}\\
\caption*{ \footnotesize  $^1$ Circled coverages correspond to correctly-specified reporting factors  }
\label{sensitivity_analysis_lag}
\end{figure}

Previous simulations demonstrate that all methods can perform well in ideal settings when reporting factors are correctly specified. However, the methods relying on reporting factor estimates can run into trouble when the reporting factors are poorly specified. We now suppose that we do not have good historical data from which to estimate these reporting factors. This may be the case, for example, if we are forecasting a newly-emerging disease. We want to evaluate how other methods (e.g., the exclusion method and the rescaling method with $\pi_{ts}(d)$ estimated using proxy shrinkage) perform for data with different amounts of under-reporting. \\
\indent We considered simulation Scenario 5, where the true reporting mechanism for each season was varied between $a=0.05$ (strong under-reporting) and $a=1$ (no under-reporting). Reporting correction, estimation, and forecasting proceeded as before. Performance diagnostics were aggregated across weeks within each season and stratified by the true value of $a$. Each result, therefore, represents an aggregation across 50 weeks in each year and across 10 simulated datasets, resulting in 500 repeated comparisons. \\
\\
\noindent \underline{Exclusion method performance as a function of $a$:} In \textbf{Supp. Figure H3}, we provide prediction errors and weighted interval scores for 1 week forecasts in 2006-2009 based on the uncorrected observed data, the validation data, and results from the exclusion method after excluding between 1 and 5 weeks of recent data. Results are shown as a function of true $a$. For the exclusion method, there is a trade-off: excluding recent data may avoid some bias due to inclusion of error-prone real-time data but will lose out on the information in the most recent data, which may also negatively impact forecast efforts. For roughly $a>0.50$, the forecast error due to throwing out information outweighed the error due to including the real-time data, and prediction error for the exclusion method was higher than uncorrected analysis. For roughly $a<0.50$, exclusion tended to outperform uncorrected analysis of the observed data, particularly in terms of weighted interval scores. These results may help guide decisions for whether or not to throw out the recent data as a function of the plausible range for $a$. \\
\\
\noindent \underline{Performance of proxy shrinkage reporting factors as a function of $a$:} \textbf{Supp. Figure H4} provides a similar exploration for the rescaling method with $\pi_{ts}(d)$ estimated using proxy shrinkage based on proxies of varying quality (i.e., correlation with $N_{ts}(\infty)$). The proxy shrinkage method was implemented defining $w_d = \frac{\omega}{d+1}$ where $\omega$ took values 0.5 or 1. When the correlation between the proxy and the validation data was very small (e.g., $<$0.15), proxy shrinkage served to increase prediction error relative to uncorrected observed data analysis. It also increased weighted interval scores relative to uncorrected analysis unless the degree of under-reporting was very high (e.g., $a <$ 0.3). When the proxy correlation was high (e.g., 0.80), rescaling using proxy shrinkage reduced prediction error relative to uncorrected analysis and reduced weighted interval scores when the degree of under-reporting was moderate to high (e.g., $a<0.7$). Any advantage of incorporating the proxy to address reporting delay depends on (1) the amount of reporting error and (2) the quality of the proxy. As a conservative rule-of-thumb, we suggest applying the proxy shrinkage method only if $a<0.5$ and the proxy is of sufficient quality (e.g., correlation $>$ 0.50). \\
\\
\noindent \underline{Sensitivity analysis with rescaling method:} In settings where the degree of reporting error may be comparatively modest but we still want to evaluate the impact of under-reporting on forecasts, a sensitivity analysis approach may be appealing. In performing a sensitivity analysis, we calculate performance metrics across different assumed values for the reporting factors. In \textbf{Figure \ref{sensitivity_analysis_lag}}, we performed such an analysis across the simulated datasets and across different \textit{true} and \textit{assumed} values for $a$ to demonstrate how coverage in 2008 and 2009 forecasts varied as a function of assumed $a$. In practice, this exploration would be conducted a single time on the observed data, where the true reporting mechanism would be unknown. Unsurprisingly, we obtained the highest coverages when the value of $a$ was correctly specified. Coverages were fairly robust to mispecification of $a$ when the true and working values of $a$ were both fairly high. When the true amount of under-reporting was very strong (e.g., a$<$ 0.10), however, even mildly misspecified $a$ resulted in a large loss of coverage. 

\section{Discussion}  \label{discuss}
Delayed reporting of infectious disease cases presents a challenge to forecasting efforts, and no general recommendations for addressing reporting delay in the forecasting setting currently exist. In this work, we synthesize many existing strategies for handling reporting delay into a single unified framework and propose a two-stage estimation procedure that can be applied to forecast modeling in general. In the first stage of the proposed estimation procedure, we leverage either (a) historical data on the accuracy of real-time reporting or (b) external data correlated with disease rates (e.g., social media or Google trends) to estimate the amount of disease under- or over-reporting. In the second stage, we describe how to implement forecast modeling, accounting for the over- or under-reporting using existing methods in the literature (sometimes, with modification). \\
\indent We applied this methodology to address reporting delay in data on influenza-like illness diagnoses in the US in 2010-2019 and data on dengue fever cases in Puerto Rico between 1990 and 2009 \citep{McGough2020}. This analysis demonstrates the potential for improving forecast performance by accounting for delayed case reporting. However, analysis of state-level influenza-like illness data highlights the need for good estimates of the current season reporting mechanisms for rescaling, mean model offset, and imputation methods; when reporting mechanisms are very poorly specified, these correction methods can sometimes do more harm than good. \\
\indent The comparative performance of the various methods was further evaluated through a simulation study. In ideal settings with correctly specified reporting mechanisms, the rescaling, mean model offset, and imputation methods resulted in reduced forecast errors relative to exclusion of the most error-prone recently reported data. This indicates that, even in the setting with very high under-reporting, the observed data can contribute useful information for forecasting when the reporting mechanism is well-understood. Caution is needed when applying the rescaling method in the setting where disease rates and anticipated inverse reporting factors are low, since this method can be sensitive to small fluctuations in the observed case data in this setting. The mean model offset and imputation methods provide additional stability in this setting and tended to result in better forecast coverage than rescaling in general. \\
\indent In implementing the rescaling, mean model offset, and imputation methods in this paper, we used a fixed estimate of $\pi_{ts}(d)$. However, this strategy does not account for our uncertainty in the estimate of $\pi_{ts}(d)$. One reasonable approach for addressing this issue is to view $\pi_{ts}(d)$ as an estimated parameter with a corresponding distribution (e.g., normal with mean $\hat \pi_{ts}(d)$ and variance based on estimate uncertainty) and apply the imputation correction method using separate \textit{draws} of $\pi_{ts}(d)$ for generating each imputation. We did not implement this approach in this paper, but we hypothesize that this approach could be useful in settings where there is a lot of unexplained variability in past reporting data.  \\
\indent Several strategies were proposed for estimating $\pi_{ts}(d)$. When past season data were available and reporting varied predictably across weeks and seasons, the lag-based and model-based estimation strategies in \ref{preest} and \ref{modelcorrection} performed well. In settings where the current season's reporting practices differed dramatically from past seasons, however, these methods failed to capture current season reporting dynamics. To address this issue, we proposed an estimator that uses the ``local" recent real-time data from the past few weeks to estimate reporting factors for the current week, providing more accurate estimates.\\
\indent In practice, historical data on reporting delay may be unavailable or incomplete and reporting may be very poorly understood. In this setting, the strategy of excluding the most recent data from forecast modeling produced more reliable forecast performance than methods that used fixed estimates of $\pi_{ts}(d)$ based on real-time data. However, this exclusion strategy involves a trade-off between loss of information in recently reported data and protection against bias caused by reporting delay. Our simulations suggest that the exclusion method is best applied when the reporting error is high (e.g., when the initial case reports are less than 50\% of the validation case counts).\\
\indent  In addition to excluding the most recent data, an alternative strategy is to leverage external data to estimate the amount of under-reporting indirectly by independently predicting the validation case counts. When these predictions are good enough, the comparison between these predicted validation case counts (here, called proxies) and the observed cases provides insight into reporting error. As with the exclusion method, this proxy approach features a trade-off between providing information for estimating reporting errors and adding more noise into the estimation. Through simulation, we evaluated how the performance of the proxy-based estimation strategy varied as a function of the amount of reporting error and the quality of the proxy (i.e., the correlation with validation case counts). We found that even poor proxies (e.g., correlation 0.13) contributed some information to under-reporting correction when the amount of under-reporting was very high. As the amount of under-reporting diminished, better and better proxies were needed to improve on uncorrected analysis.\\
\indent Synthesizing the insights from our data analysis and simulations, we produced some general recommendations for the handling of reporting delay in \textbf{Figure \ref{methodsdiagram}b}. These recommendations provide guidance for handling delay in disease case reporting and may serve as a useful resource to inform practical infectious disease forecasting efforts. Overall, our results clearly demonstrate potential benefits of accounting for reporting delay to improve forecast performance. However, the best-performing methods require either high-quality data on historical reporting mechanisms or strongly predictive proxies of validation case counts. Historical reporting data are often not recorded or not publicly available for many commonly-studied diseases. This motivates additional thinking about disease reporting infrastructure, particularly as new initiatives are underway to expand reporting and forecasting capabilities at the US national level \citep{US2021}. \\
\indent This work focuses on the problem of handling reporting delay and defines the ``true" case rates we want to predict as the eventually-reported validation case rates. In reality, many diseases may be poorly or incompletely captured by official reporting mechanisms \citep{Gibbons2014}, as plainly demonstrated by the COVID-19 pandemic. Given estimates of the amount of reporting error relative to the unobserved ``true" case counts, the rescaling, mean model offset, and imputation methods discussed in this paper can be applied. The challenge is then to estimate the amount of reporting error, a discussion of which is beyond the scope of this work. Alternatively, this problem can be handled in a sensitivity analysis framework, where we can compare the robustness of our forecasts across multiple plausible assumptions about the amount of reporting error relative to the ``true" case counts. Additional work is needed to leverage multiple data streams to inform estimation of reporting error when the target is defined as the never-observed ``true" case counts.

\section*{Acknowledgments}
\indent Dr. Beesley was funded by Los Alamos National Laboratory LDRD 20210761PRD1. Dr. Osthus and Dr. Del Valle were funded by NIH/NIGMS under grant R01GM130668-01. This  work  is  approved  for  distribution  under LA-UR-21-30640. The findings and conclusions in this report are those of the authors and do not necessarily represent the official position of Los Alamos National Laboratory. Los Alamos National Laboratory, an affirmative action/equal opportunity employer, is managed by Triad National Security, LLC, for the National Nuclear Security Administration of the U.S. Department of Energy under contract 89233218CNA000001. The funders had no role in study design, data collection and analysis, decision to publish, or preparation of the manuscript.

\section*{Data Availability Statement}
All data are publicly available. ILInet data are available through the DELPHI API at \url{https://github.com/cmu-delphi/delphi-epidata}. Puerto Rico dengue fever data are provided as part of R package \textit{NobBS}. Example code will be provided on GitHub when approved for distribution. 

\newpage
\bibliographystyle{unsrtnat}
\bibliography{Bib}

\end{document}